\documentclass{elsart}

\usepackage{amsmath}
\usepackage{citesort}
\usepackage{epic}
\usepackage{eepic}
\usepackage{epsfig}
\usepackage{epic}
\usepackage{eepic}

\begin{document}
\newcommand{\secref}[1]{section \ref{#1}}
\newcommand{\appref}[1]{appendix \ref{#1}}
\newcommand{\figref}[1]{figure \ref{#1}}
\newcommand{\tabref}[1]{table \ref{#1}}
\newcommand{\Secref}[1]{Section \ref{#1}}
\newcommand{\Appref}[1]{Appendix \ref{#1}}
\newcommand{\Figref}[1]{Figure \ref{#1}}
\newcommand{\Tabref}[1]{Table \ref{#1}}
\newcommand{\ie}{i.e.\ }
\newcommand{\vm}{\ensuremath v_{max}}
\newcommand{\closediv}[1]{\ensuremath \lbrack #1 \rbrack}
\newcommand{\theset}[1]{\ensuremath \lbrace #1 \rbrace}

\newcommand{\onefig}[4]{
  \begin{figure}[ht]
    \centering{\epsfig{figure=#1,height=#2}}
    \caption{#3}
    \label{#4}
  \end{figure}
  }
\newcommand{\twofigs}[5]{
  \begin{figure}[ht]
    \centerline{\epsfig{figure=#1,height=#3}
      \epsfig{figure=#2,height=#3}}
    \caption{#4}
    \label{#5}
  \end{figure}
  }


\runauthor{Burstedde et al.}
\begin{frontmatter}
\title{Simulation of pedestrian dynamics using a 2-dimensional cellular
  automaton}
\author[Cologne]{C.~Burstedde\thanksref{EmailCB}}
\author[Cologne]{K.~Klauck\thanksref{EmailKK}}
\author[Cologne]{A.~Schadschneider\thanksref{EmailAS}}
\author[Cologne]{J.~Zittartz\thanksref{EmailJZ}}
\address[Cologne]{Institut f\"ur Theoretische Physik, Universit\"at zu 
K\"oln, D-50923 K\"oln, Germany}
\thanks[EmailCB]{E-mail: cb@thp.uni-koeln.de}
\thanks[EmailKK]{E-mail: kok@thp.uni-koeln.de}
\thanks[EmailAS]{E-mail: as@thp.uni-koeln.de}
\thanks[EmailJZ]{E-mail: zitt@thp.uni-koeln.de}

\begin{abstract}
  We propose a 2-dimensional cellular automaton model to simulate
  pedestrian traffic.  It is a $\vm = 1$ model with exclusion
  statistics and parallel dynamics.  
  Long-range interactions between the pedestrians are mediated by a 
  so called \emph{floor field} which modifies the transition rates to
  neighbouring cells. This field, which can be discrete or continuous,
  is subject to diffusion and decay. Furthermore it can be modified by
  the motion of the pedestrians. Therefore the model uses an idea 
  similar to chemotaxis, but with pedestrians following a virtual
  rather than a chemical trace.  Our main goal is to
  show that the introduction of such a floor field is sufficient to model
  collective effects and self-organization encountered in pedestrian
  dynamics, e.g.\ lane formation in counterflow through a large corridor.
  As an application we also present simulations of the evacuation of a
  large room with reduced visibility, e.g.\ due to failure of lights or smoke.
\end{abstract}

\vspace{0.7cm}
\end{frontmatter}



\section{Introduction}

Considerable research has been done on the topic of traffic flow 
using methods from physics during the last decade 
\cite{juelich,tgf97,tgf99,helb,chowd,review2,nagel99,dhrev}.
Cellular automata inspired by the pioneering works
\cite{NagelS,Ito,BML} compose by now an important class of models.  
Most studies have been devoted to one-dimensional systems, where 
several analytic approaches exist to calculate or approximate the 
stationary state.

The majority of these models deals with particles which can move by
more than one cell per time step (maximal velocity $\vm > 1$).  Furthermore,
it seems to be widely accepted that the most suitable update procedure is the
parallel (synchronous) update.  Both open and periodic boundary conditions 
have been considered, where problems with open boundaries are generally 
harder to treat analytically (for a review, see \cite{derrida}).

On the other hand, pedestrian dynamics has not been studied as extensively
as vehicular traffic, especially using a cellular automata approach.
One reason is probably its generically two-dimensional nature.
In recent years, continuum models have been most successful in modelling
pedestrian dynamics. An important example are the {\em social force models}
(see e.g.\ \cite{helb,dhrev,social} and references therein). 
Here pedestrians are treated as particles subject to
long-ranged\footnote{typically decaying exponentially} forces
induced by the social behaviour of the individuals. This leads to
(coupled) equations of motion similar to Newtonian mechanics. There are,
however, important differences since e.g.\ in general the third law
(``actio = reactio'') is not fulfilled.

In contrast to the social force models our approach is closer in spirit
to the general strategy of modelling (elementary) forces on a microscopic
level by the exchange of mediating particles which are bosons. It is 
therefore similar to {\em active walker models} \cite{activewalker,trail}
used so far mainly to describe trail formation, chemotaxis 
(see \cite{benjacob} for a review) etc. Here
the walker leaves a trace by modifying the underground on his path.
This modification is real in the sense that it could be measured in
principle. For trail formation vegetation is destroyed by the walker
and in chemotaxis he leaves a chemical trace. In contrast in our model
the trace is virtual. Its main purpose is to transform effects of
long-ranged interactions (e.g.\ following people walking some distance
ahead) into a local interaction (with the ``trace''). This allows e.g.\
for a much more efficient simulation on a computer.

Cellular automata for pedestrian dynamics have been proposed in
\cite{fukui1,fukui2,nagatani1,nagatani2,nagatani3,hubert}.
These models can be considered as generalizations of the 
Biham-Middleton-Levine model for city traffic \cite{BML}.
Most works have focussed on the occurrence of a jamming transition
as the density of pedestrians is increased. 
All models have $\vm=1$, except for the generalization proposed in
\cite{hubert} which is used for analyzing evacuation processes
on-board passenger ships. The other models use a kind of 
"sublattice-dynamics" which distinguishes between different types
of pedestrians according to their preferred walking direction.
Such an update is not easy to generalize to more complex situations
where the walking direction can change.
To our knowledge so
far most other collective effects encountered empirically 
\cite{helb,CrowdFluids,CrowdFluids2,weidmann,panic} have not been
reproduced using these models.
Another discrete model has been proposed earlier by Gipps and 
Marksj\"os \cite{gipps}. This model is somewhat closer in spirit to 
our model than the cellular automata approaches of 
\cite{fukui1,fukui2,nagatani1,nagatani2,nagatani3,hubert} since the
transitions are determined by the occupancies of the neighbouring cells.
However, also this model can not reproduce all the collective effects.
In \cite{bolay} a discretized version of the social force model
has been introduced. The repulsive potentials by the pedestrians are
stored in a global potential, with pedestrians reacting to the
gradients of this global potential. Although this model is able
to reproduce collective effects it suffers from some drawbacks
\cite{privcomm}. It is not flexible enough to treat individual reactions
to other pedestrians, and collision-avoidance is not always guaranteed
for velocities greater than 1.

First we discuss some general principles we took into account in the
development of our model. In contrast to vehicular traffic the time 
needed for acceleration and braking is negligible.
The velocity distribution of pedestrians is sharply peaked
\cite{CrowdFluids}.  These facts naturally lead to a model with 
$\vm =1$ (if space is discrete), i.e.\ only transitions to nearest
neighbours\footnote{Here the four diagonal neighbours are included.}
are allowed.  
Furthermore, a greater $\vm$ would be harder to implement in 2 dimensions,
especially when combined with parallel dynamics,
and reduce the computational efficiency.
The number of possible target cells increases quadratically with the 
interaction range. Furthermore one has to check whether the path
is blocked by other pedestrians. This might even be ambigious
for diagonal motion and crossing trajectories.

To keep the model simple, we strongly emphasize the principle to
provide the particles with as little intelligence as possible and to
achieve the formation of complex structures and collective effects by
means of self-organization.  Effectively, there is absolutely no
intelligence (\ie look-ahead distances or multiple moves per
update step depending on the distribution of occupied neighbour sites)
in our model. In contrast to older approaches we do not make 
detailed assumptions about the human behaviour. Nevertheless
the model is able to reproduce many of the basic phenomena.

The key feature to substitute individual intelligence is the floor
field.  Apart from the occupation number each cell carries an additional
quantity (field) which can be either discrete or continuous.  
This field can have its own dynamics given by 
diffusion and decay coefficients.  

Interactions between pedestrians are repulsive for short distances.
One likes to keep a minimal distance to others in order to avoid
bumping into them. In the simplest version of our model this is taken
into account through hard-core repulsion which prevents multiple
occupation of the cells. For longer distances the interaction is
often attractive. E.g.\ when walking in a crowded area it is usually
advantageous to follow directly behind the predecessor. Large crowds
may also be attractive due to curiosity.

In order to produce a flow around obstacles in a simple way we also
present a variant of our model where the pedestrians can be in one
of two modes (or moods), "happy" or "unhappy". These two modes are
distinguished by their fluctuations. "Happy" pedestrians try to
move in a preferred direction whereas "unhappy" pedestrians move in
a more random fashion. 
This is sufficient to avoid a jamming transition due to obstacles at
unrealistical low densities.  

With two particle species moving in opposite 
directions, each with its own
floor field, effects can be observed which are so far only achieved by
continuous models: lane formation and oscillation of the direction of
flow at doors.  We consider this model to be another proof of the
ability of cellular automata to create complex behaviour out of simple
rules and the great applicability to all kinds of traffic flow problems.

The model can be used together with models for route selection which
assign certain routes to each pedestrian. It
only assumes that at every timestep for each pedestrian a transition
matrix (matrix of preferences) is given.


\section{Model}

The underlying structure is a 2-dimensional grid which can be closed
periodically in one or both directions.  
Each cell can either be empty or occupied by exactly one particle
(pedestrian). The size of a cell corresponds to 
approximately $40\times 40~cm^2$. This is the typical space occupied
by a pedestrian in a dense crowd \cite{weidmann}.
For special situations it might be desirable to use a finer discretization,
e.g.\ such that each pedestrian occupies four cells instead of one. In
this paper, however, we concentrate on the simplest case which seems to be
sufficient for most purposes.
The update is done in parallel for all particles. This introduces
a timescale into the dynamics which can roughly be identified with the
reaction time $t_{\rm reac}$. In the deterministic limit, corresponding
to the maximal possible walking velocity in our model, a single
pedestrian (not interacting with others) moves with a velocity of
one cell per timestep, i.e.\ $40~cm$ per timestep. Empirically the
average velocity of a pedestrian is about $1.3~m/s$ \cite{weidmann}.
This gives an estimate for the real time corresponding to one timestep
in our model. It is approximately $0.3~sec$, i.e.\ of the order of
the reaction time $t_{\rm reac}$, and thus is consistent with our
microscopic rules.


\subsection{Basic Rules}
\label{sec_rules}

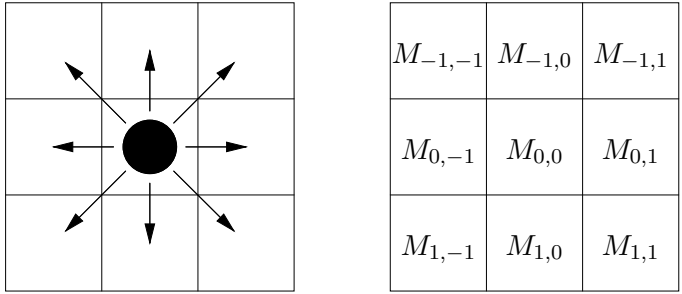
\begin{figure}
  \begin{center}
    \setlength{\unitlength}{0.00083333in}
\begingroup\makeatletter\ifx\SetFigFont\undefined%
\gdef\SetFigFont#1#2#3#4#5{%
  \reset@font\fontsize{#1}{#2pt}%
  \fontfamily{#3}\fontseries{#4}\fontshape{#5}%
  \selectfont}%
\fi\endgroup%
{\renewcommand{\dashlinestretch}{30}
\begin{picture}(4120,1839)(0,-10)
\path(1212,1812)(1212,12)
\path(612,1812)(612,12)
\path(12,612)(1812,612)
\path(12,1812)(1812,1812)(1812,12)
	(12,12)(12,1812)
\path(12,1212)(1812,1212)
\path(3612,1812)(3612,12)
\path(3012,1812)(3012,12)
\path(2412,612)(4212,612)
\path(2412,1812)(4212,1812)(4212,12)
	(2412,12)(2412,1812)
\path(2412,1212)(4212,1212)
\put(912,912){\blacken\ellipse{336}{336}}
\put(912,912){\ellipse{336}{336}}
\path(1137,912)(1512,912)
\path(1137,912)(1512,912)
\blacken\path(1392.000,882.000)(1512.000,912.000)(1392.000,942.000)(1392.000,882.000)
\path(912,687)(912,312)
\path(912,687)(912,312)
\blacken\path(882.000,432.000)(912.000,312.000)(942.000,432.000)(882.000,432.000)
\path(687,912)(312,912)
\path(687,912)(312,912)
\blacken\path(432.000,942.000)(312.000,912.000)(432.000,882.000)(432.000,942.000)
\path(912,1137)(912,1512)
\path(912,1137)(912,1512)
\blacken\path(942.000,1392.000)(912.000,1512.000)(882.000,1392.000)(942.000,1392.000)
\path(1062,1062)(1437,1437)
\path(1062,1062)(1437,1437)
\blacken\path(1373.360,1330.934)(1437.000,1437.000)(1330.934,1373.360)(1373.360,1330.934)
\path(1062,762)(1437,387)
\path(1062,762)(1437,387)
\blacken\path(1330.934,450.640)(1437.000,387.000)(1373.360,493.066)(1330.934,450.640)
\path(762,762)(387,387)
\path(762,762)(387,387)
\blacken\path(450.640,493.066)(387.000,387.000)(493.066,450.640)(450.640,493.066)
\path(762,1062)(387,1437)
\path(762,1062)(387,1437)
\blacken\path(493.066,1373.360)(387.000,1437.000)(450.640,1330.934)(493.066,1373.360)
\put(2712,1437){\makebox(0,0)[b]{\smash{{{\SetFigFont{11}{13.2}{\familydefault}{\mddefault}{\updefault}$M_{-1,-1}$}}}}}
\put(3312,1437){\makebox(0,0)[b]{\smash{{{\SetFigFont{11}{13.2}{\familydefault}{\mddefault}{\updefault}$M_{-1,0}$}}}}}
\put(3912,837){\makebox(0,0)[b]{\smash{{{\SetFigFont{11}{13.2}{\familydefault}{\mddefault}{\updefault}$M_{0,1}$}}}}}
\put(3312,837){\makebox(0,0)[b]{\smash{{{\SetFigFont{11}{13.2}{\familydefault}{\mddefault}{\updefault}$M_{0,0}$}}}}}
\put(2712,837){\makebox(0,0)[b]{\smash{{{\SetFigFont{11}{13.2}{\familydefault}{\mddefault}{\updefault}$M_{0,-1}$}}}}}
\put(2712,237){\makebox(0,0)[b]{\smash{{{\SetFigFont{11}{13.2}{\familydefault}{\mddefault}{\updefault}$M_{1,-1}$}}}}}
\put(3312,237){\makebox(0,0)[b]{\smash{{{\SetFigFont{11}{13.2}{\familydefault}{\mddefault}{\updefault}$M_{1,0}$}}}}}
\put(3912,237){\makebox(0,0)[b]{\smash{{{\SetFigFont{11}{13.2}{\familydefault}{\mddefault}{\updefault}$M_{1,1}$}}}}}
\put(3912,1437){\makebox(0,0)[b]{\smash{{{\SetFigFont{11}{13.2}{\familydefault}{\mddefault}{\updefault}$M_{-1,1}$}}}}}
\end{picture}
}
    \caption{A particle, its possible transitions and the 
associated matrix of preference $M=(M_{ij})$.}
\label{fig_prefs}
  \end{center}
\end{figure}

Each particle is given a direction of preference.  
>From this direction, a $3 \times 3$ \emph{matrix of preferences} is
constructed which contains the probabilities for a move of the
particle. The central element describes the probability for the
particle not to move at all, the remaining 8 correspond to a move to
the neighbouring cells.
The probabilities can be related to the velocity and the
longitudinal and transversal standard deviations 
(see \appref{MoD} for details).  
In practice all particles of the same species share the
values of these parameters and in consequence the same matrix.  In the
simplest case the pedestrian is allowed to move in one direction only 
without fluctuations and in the corresponding matrix of preference only one
element is one and all others are zero (see Fig.~\ref{fig_prefs}).

This ansatz can easily be extended by fixing the direction of
preference for each cell separately, e.g.\ to handle structures inside
buildings.  Then the particles would use the matrix
belonging to the cell they occupy at a given step.

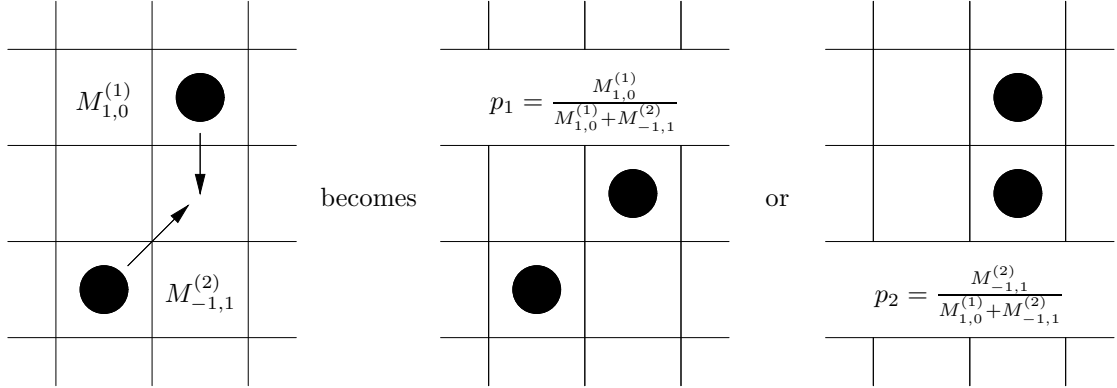
\begin{figure}
  \begin{center}
    \setlength{\unitlength}{0.00083333in}
\begingroup\makeatletter\ifx\SetFigFont\undefined%
\gdef\SetFigFont#1#2#3#4#5{%
  \reset@font\fontsize{#1}{#2pt}%
  \fontfamily{#3}\fontseries{#4}\fontshape{#5}%
  \selectfont}%
\fi\endgroup%
{\renewcommand{\dashlinestretch}{30}
\begin{picture}(8173,2439)(0,-10)
\put(1212,1812){\blacken\ellipse{300}{300}}
\put(1212,1812){\ellipse{300}{300}}
\put(612,612){\blacken\ellipse{300}{300}}
\put(612,612){\ellipse{300}{300}}
\put(3912,1212){\blacken\ellipse{300}{300}}
\put(3912,1212){\ellipse{300}{300}}
\put(3312,612){\blacken\ellipse{300}{300}}
\put(3312,612){\ellipse{300}{300}}
\put(6312,1812){\blacken\ellipse{300}{300}}
\put(6312,1812){\ellipse{300}{300}}
\put(6312,1212){\blacken\ellipse{300}{300}}
\put(6312,1212){\ellipse{300}{300}}
\path(1212,1587)(1212,1212)
\path(1212,1587)(1212,1212)
\blacken\path(1182.000,1332.000)(1212.000,1212.000)(1242.000,1332.000)(1182.000,1332.000)
\path(762,762)(1137,1137)
\path(762,762)(1137,1137)
\blacken\path(1073.360,1030.934)(1137.000,1137.000)(1030.934,1073.360)(1073.360,1030.934)
\path(312,2412)(312,12)
\path(912,2412)(912,12)
\path(1512,2412)(1512,12)
\path(12,2112)(1812,2112)
\path(1812,1512)(12,1512)
\path(12,912)(1812,912)
\path(1812,312)(12,312)
\path(2712,2112)(4512,2112)
\path(4512,1512)(2712,1512)
\path(2712,912)(4512,912)
\path(4512,312)(2712,312)
\path(3612,2412)(3612,2112)
\path(3612,2412)(3612,2112)
\path(3612,1512)(3612,12)
\path(3612,1512)(3612,12)
\path(3012,2412)(3012,2112)
\path(3012,2412)(3012,2112)
\path(4212,2412)(4212,2112)
\path(4212,2412)(4212,2112)
\path(4212,1512)(4212,12)
\path(4212,1512)(4212,12)
\path(3012,1512)(3012,12)
\path(3012,1512)(3012,12)
\path(5112,2112)(6912,2112)
\path(6912,1512)(5112,1512)
\path(5112,912)(6912,912)
\path(6912,312)(5112,312)
\path(6012,2412)(6012,912)
\path(6012,2412)(6012,912)
\path(6012,312)(6012,12)
\path(6012,312)(6012,12)
\path(6612,12)(6612,312)
\path(6612,12)(6612,312)
\path(5412,12)(5412,312)
\path(5412,12)(5412,312)
\path(5412,912)(5412,2412)
\path(5412,912)(5412,2412)
\path(6612,2412)(6612,912)
\path(6612,2412)(6612,912)
\put(1212,537){\makebox(0,0)[b]{\smash{{{\SetFigFont{10}{12.0}{\familydefault}{\mddefault}{\updefault}$M_{-1,1}^{(2)}$}}}}}
\put(3612,1737){\makebox(0,0)[b]{\smash{{{\SetFigFont{10}{12.0}{\familydefault}{\mddefault}{\updefault}$p_1 = \frac{M_{1,0}^{(1)}}{M_{1,0}^{(1)}+M_{-1,1}^{(2)}}$}}}}}
\put(6012,537){\makebox(0,0)[b]{\smash{{{\SetFigFont{10}{12.0}{\familydefault}{\mddefault}{\updefault}$p_2 = \frac{M_{-1,1}^{(2)}}{M_{1,0}^{(1)}+M_{-1,1}^{(2)}}$}}}}}
\put(2262,1137){\makebox(0,0)[b]{\smash{{{\SetFigFont{10}{12.0}{\familydefault}{\mddefault}{\updefault}becomes}}}}}
\put(4812,1137){\makebox(0,0)[b]{\smash{{{\SetFigFont{10}{12.0}{\familydefault}{\mddefault}{\updefault}or}}}}}
\put(612,1737){\makebox(0,0)[b]{\smash{{{\SetFigFont{10}{12.0}{\familydefault}{\mddefault}{\updefault}$M_{1,0}^{(1)}$}}}}}
\end{picture}
}
    \caption{Solving conflicts according to the relative probabilities for
the case of two particles with matrices of preference $M^{(1)}$ and
$M^{(2)}$.}
\label{fig_conflict}
  \end{center}
\end{figure}

In each update step, for each particle a desired move is chosen
according to these probabilities. This is done in parallel for
all particles. If the target cell is occupied, the
particle does not move.  If it is not occupied and no other particle
targets the same cell, the move is executed.  If more than one
particle share the same target cell, one is chosen according to the
relative probabilities with which each particle chose their target.
This particle moves while its rivals for the same target keep their
position (see Fig.~\ref{fig_conflict}).

The matrix used to achieve maximum flow is clearly the simplest case
of unidirectional fluctuation-free motion described above.  
Transversal fluctuations reduce the flow by
introducing interference between lanes.  However, this setting is not
sufficient to avoid e.g.\ jams behind obstacles.  To escape a jammed
situation, the particles need a mode where they can select between
different cells to move backwards or sideways in order to
eventually make their way around the obstacle.

The rules presented up to here are a straightforward generalization of the
CA rules used so far for the description of traffic flow
\cite{fukui1,fukui2,nagatani1,nagatani2,nagatani3}.
The main difference is that in principle transitions in all directions
are possible and each pedestrian $j$ might have her own preferred
direction of motion characterized by a matrix of preferences $M^{(j)}$.
The only interaction between particles taken into account so far is 
hard-core exclusion.


\subsection{Floor Field}
\label{sec_floor}

In order to reproduce certain collective phenomena it is necessary to
introduce further longer-ranged interactions. In some continuous models
this is done using the idea of a social force \cite{helb,dhrev,social}.
Here we introduce a different approach. 
Since we want to keep the model as simple as possible we try to avoid
using a long-range interaction explicitly. Instead we introduce the concept
of a {\em floor field} which is modified by the pedestrians and which in turn
modifies the transition probabilities. This allows to take into
account interactions between pedestrians and the geometry of the system
(building) in a unified and simple way without loosing the advantages 
of local transition rules. The floor field modifies the transition
probabilities in such a way that a motion into the direction of larger
fields is preferred. 

The floor field can be thought of as a second grid of cells underlying
the grid of cells occupied by the pedestrians. It can be discrete or
continuous.  In this paper we will give examples for both variants.
In general we distinguish between static and dynamic floor fields. The
{\em static floor field} $S$ does not evolve with time and is not
changed by the presence of pedestrians. Such a field can be used to
specify regions of space which are more attractive, e.g.\ an emergency
exit (see the example in Sec.~\ref{evacuation}) or shop windows. This
has an effect similar to a position-dependent matrix of preference but
is much easier to realize.

In contrast the {\em dynamic floor field} $D$ is modified by the presence
of pedestrians and has its own dynamics, i.e.\ diffusion and decay.
Usually the dynamic floor field is used to model a (``long-ranged'')
attractive interaction
between the pedestrians. Each pedestrian leaves a ``trace'', i.e.\ the
floor field of occupied cells is increased. Since the total transition
probability is proportional to the dynamic floor field it becomes more
attractive to follow in the footsteps of other pedestrians. Explicit 
examples where such an interaction is relevant will be given in 
Sec.~\ref{sec_simus}. The dynamic floor field is also subject to diffusion
and decay which leads to a dilution and finally the vanishing of the
trace after some time.

In general the {\em transition probability} $p_{ij}$ in direction $(i,j)$
(see Fig.~\ref{fig_prefs}) is given by\footnote{Note that this is not
a product of matrices but just the product of the
corresponding matrix elements.}
\begin{equation}
p_{ij}=NM_{ij}D_{ij}S_{ij}(1-n_{ij}).
\label{transprob}
\end{equation}
Here $n_{ij}$ is the occupation number of the target cell in direction
$(i,j)$, i.e.\ $n_{ij}=0$ for an empty cell and $n_{ij}=1$ for an 
occupied cell. Therefore transitons to occupied cells are forbidden.
$N$ is a normalization factor to ensure $\sum_{(i,j)}p_{ij}=1$
where the sum is over the nine possible target cells.
In Sec.~\ref{sec_desire} we will also use a slightly different
form for the transition probabilities which is more general than
(\ref{transprob}).

The update rules of the full model including the interaction with the
floor fields then have the following structure:
\begin{enumerate}
\item The dynamic floor field $D$ is modified according to its diffusion 
and decay rules (see Sec.~\ref{discfloor} and \ref{contfloor}).
\item For each pedestrian, the transition probabilities for a move to an
unoccupied neighbour cell $(i,j)$ is determined by the matrix of preferences 
and the local dynamic and static floor fields, e.g. $p_{ij}\propto
M_{ij}D_{ij}S_{ij}$.
\item Each pedestrian chooses a target cell based on the probabilities
of the transition matrix $P=(p_{ij})$.
\item The conflicts arising by any two or more pedestrians attempting to
  move to the same target cell are resolved, e.g.\ using the procedure
described in Sec.~\ref{sec_rules}.
\item The pedestrians which are allowed to move execute their step.
\item The pedestrians alter the dynamic floor field of the cell they occupied
  before the move.
\end{enumerate}
The explicit form of the interaction between the pedestrians and the floor
field and the dynamics of the floor field will be specified
in Sec.~\ref{discfloor} and \ref{contfloor}.

If more than one pedestrian species exists (\ie two groups moving in
opposite directions), each species interacts with its own floor field.
In the simplest case these fields are independent from each other.


\subsection{Discrete floor fields}
\label{discfloor}

In the discrete case the fields are realized through noninteracting 
particles which do not obey a hard-core exclusion principle. Therefore
they will be called {\em bosons} in the following. Since the particles
corresponding to the pedestrians are not allowed to share a cell these
will be called {\em fermions}.
The fermions couple to the bosons locally which drives the
fermions in a preferred direction and induces a long-range interaction
between the fermions. 

The essence of our approach is that for each fermion the probability
to jump into a direction with a larger number of bosons is
increased. Thus, the motion is simply driven by gradients in the 
floor field, i.e.\ in the density of the bosons. 

The first type of bosons ($s$-bosons) is completely static. At the
beginning of a simulation for each cell $(x,y)$ the occupation
number of $s$-bosons $\tau_s(x,y)$ is fixed to a specific
value. Furthermore, at the beginning every cell is void of
bosons of the second type, the dynamic bosons ($d$-bosons). 
Whenever a fermion jumps from site $(x,y)$ to one of the neighbouring 
cells, the $d$-boson occupation number of cell $(x,y)$ is increased 
by one (fermions leave a trace):
\begin{equation}
\tau_d(x,y)\rightarrow \tau_d(x,y) +1.
\end{equation}  
After all motions of the fermions during one timestep have been
performed, the oldest $d$-boson of each cell
is destroyed with probability $\alpha$, if the
lifetime of this boson is larger than one (i.e.~it has been created
during the previous update step or earlier).

Now that the two bosonic floor fields have been introduced, the
update procedure for the fermions can be given. At every discrete time
step $t\rightarrow t+1$ each fermion verifies which of its neighbouring
cells $(i,j)$ are empty ($n_{ij}=0$). 
The transition probability to occupied neighbouring cells is
set to zero. Thus, the probability for a jump from the
center cell $(0,0)$ to an unoccupied neighbour site $(i,j)$ is given by
\begin{equation}
{p}_{ij}=N\exp(\beta J_s \triangle_s(i,j))\cdot\exp(\beta J_d
\triangle_d(i,j))\cdot (1-n_{ij})\cdot d_{ij}
\end{equation} 
where
\begin{equation}
\triangle_s(i,j)=\tau_s(i,j)-\tau_s(0,0)\qquad \text{and}\qquad 
\triangle_d(i,j)=\tau_d(i,j)-\tau_d(0,0).
\end{equation}
$N$ is again a normalization factor to ensure $\sum_{(i,j)}p_{ij}=1$.
$d_i$ is a correction factor taking into account the direction the
particle in cell  $0$ has been coming from. The variables $J_s$ and $J_d$
control the coupling strength between the fermions and the
$s$-bosons and the $d$-bosons, respectively. $\beta$ plays
the r\^{o}le of an inverse temperature.  Note that the $d$-bosons lead
to a long range interaction between fermions in space and time. 

The correction factor is introduced in order to prevent that the
fermions are not confused by their own trace. One has to
distinguish between three cases: If the fermion at $0$ has been
sitting at $i$ at time $t-1$, the $d$-boson sitting on top of $i$ has
been left by the fermion under consideration. Setting
\begin{equation}
d_{ij}=\exp(- \beta J_d)
\end{equation}  
this fermion is not taken into account in the calculation of the
transition probabilities.
If during the time step from $t-1\rightarrow t$ the fermion has moved
into the  direction of the vector pointing from  $(0,0)\rightarrow (i,j)$,
 this motion shall be enhanced
\begin{equation}
d_{ij}=\exp(\beta J_0).
\end{equation}
Therefore, $J_0$ is a parameter which can be used to tune the inertia of
the fermions. In all other case $d_{ij}$ equals one.

This prescription is not free of collisions. Therefore, if
$m$ fermions try to perform a move onto the same site,
only  one of these fermions is allowed to perform this move. This
fermion is picked at random with probability $1/m$. 
Of course one can also use the method described in Sec.~\ref{sec_rules}
for the resolution of the conflicts. For the problem studied in 
Sec.~\ref{evacuation} the details of the conflict resolution turned out
to play no important role and we therefore used the simpler rule.

\subsection{Continuous floor field}
\label{contfloor}

In the continuous variant each cell $j$ of the floor field 
carries a continuous field value $f_j$ between 0 and 1. 
The basic purpose of the floor field is again to determine the transition
probabilities of the pedestrians.

In the example studied in Sec.~\ref{laneformation}
we will only use a dynamic floor field, but a
generalization which includes also a static field is straightforward.
Since we are interested in applications related to the flow around
obstacles we introduce two kinds of states in which pedestrian can be:
"happy" or "unhappy".
A pedestrian becomes "unhappy" if several consecutive desired moves
could not be carried out due to conflicts. She then changes her
strategy, i.e.\ her matrix of preferences (see Sec.~\ref{sec_desire}).
The interaction of the pedestrians with the floor field is then as follows: 
"Happy" pedestrians locally increase the field, and a
large field aids "unhappy" pedestrians to become "happy" again, in a
sense to be specified below.
This is sufficient to produce a flow around obstacles, e.g.\
lane formation (see Sec~\ref{laneformation}) or oscillations of the 
direction of flow at doors. 

Without the distinction between the two states pedestrians would have
the tendency to pile up in front of obstacles. If a constant flow from
behind exists it will become increasingly difficult for these
pedestrians to turn around and avoid the obstacle and therefore the
pile will grow. For static obstacles this effect is related to the 
fact that our pedestrians have minimal intelligence. This assumption
should be well justified in situations where they move in unknown
territory and at a reduced visibility, e.g.\ due to smoke or failing
lights. In normal situations the pedestrians can see a static obstacle
(e.g.\ a wall) from some distance and will try to avoid them early.
This can easily be incorporated in our model using a static floor
field which becomes smaller just in front of the obstacle and
thus reduces the corresponding transition probabilities.
For dynamical obstacles, e.g.\ other pedestrians moving in the 
opposite direction, one way of avoiding unrealistic jamming properties
is the introduction of different modes.


\subsubsection{Diffusion and decay}

The dynamic floor field $F$ is subject to diffusion and decay.
It evolves according to 
\begin{equation}
  \frac{\partial F}{\partial t} = D \cdot \Delta F - \delta \cdot F
\label{eq_diffu}
\end{equation}
which is discretized in the standard manner. Here $D$ is the diffusion 
constant and $\delta$ the decay constant.
The ranges are restricted to $D \in \closediv {0, \frac{1}{8}}$ and
$\delta \in \closediv {0,\frac{1}{2}}$ to suppress oscillations in the 
floor fields and insure that the values do not leave the
interval $[0,1]$.

\subsubsection{Floor field affects the pedestrian's desire and state}
\label{sec_desire}

As mentioned above there are different ways to model the interaction of 
the floor fields with the pedestrians. Here we present a preliminary 
solution for the situation of pedestrians with minimal intelligence
which can probably be simplified further. The form of the transition rates
is slightly more general than (\ref{transprob}). Furthermore we
allow the pedestrians to be in two different modes (or moods) described
by different matrices of preference.
The transitions between these two modes are controlled by the floor
field. The rates are, however, subject to two restrictions:
\begin{itemize}
\item A uniform floor field should not alter the matrix of preferences.
\item A non-uniform floor field should be able to change a matrix
  element from zero to a nonzero value.
\end{itemize}

A per-element addition of the matrix of preferences and the floor field
violates the first principle, while a multiplication by e.g.\ Boltzmann
factors violates the second.  Therefore, we are working with a
compromise between the two by slightly generalizing eq.~(\ref{transprob})
\begin{equation}
  p_{ij} = N\left( M_{ij} + b_2 \right) \cdot
  \exp \big( \left( F_{ij} - F_{avg} \right) \cdot b_1 \big) \, \text{,}
\label{eq_pij}
\end{equation}
where $F$ denotes the floor field matrix, $F_{avg}$ the floor field
averaged over all nine relevant cells and $P=(p_{ij})$ the transition matrix
which has to be normalized by a normalization factor $N$.  
The parameters of this rule are $b_1$ and $b_2$.

We introduce a second mode by switching to a different
matrix of preferences.  The mode described so far will
be called \emph{happy}, while the \emph{unhappy} mode is realized by a
different matrix of preferences which is simply characterized by greater
standard deviations and a reduced velocity. This happens typically in
high-density situations. The motion then becomes less directed and the
larger fluctuations help to avoid clogging. Happy pedestrians which
could not move to their desired target field in several consecutive
timesteps (the exact number is a parameter of the model, in our simulations 
set to 3) enter the unhappy mode, while unhappy pedestrians become happy
again after a certain number of consecutive allowed desires (4 in our
simulations).  The optimal values for these two parameters depend on the
preferences of the given simulation (\ie maximum flow versus
flexible obstacle avoidance).  Of course this requires a minimal
per-pedestrian memory consisting of two counters.

One could easily alter this definition by allowing a continuous
choice of matrix by interpolating between the two states.  
This would correspond to introducing a continuous \emph{spectrum of moods}
instead of the discrete states "happy" and "unhappy".

In addition to the mechanism for transitions between the modes
described above, a unhappy pedestrian changes to a happy one immediately 
if the value of
the floor field at its cell of origin is greater than a certain threshold.  
This leads to a smooth integration into a calm
region of high flow of pedestrians which have run into an obstacle or a
jam.

\subsubsection{Pedestrians affect the floor field}

After a pedestrian has completed a certain number of total allowed
moves  (the value of which is usually 3) in the happy mode, the value of 
the floor field of its originating cell is increased.  
This introduces a third counter residing in the per-pedestrian memory.  
The counter is set to zero at mode changes.  The prescription to alter 
the floor field reads
\begin{equation}
  F \rightarrow F + \min \big( \left(1 - F\right) g_1, \, g_2 \big).
\label{eq_g1g2}
\end{equation}
The parameters of this rule are $g_1 \in \closediv{0, 1}$ and $g_2 \in
\closediv{0,1}$.  The way how pedestrians affect the floor field
can certainly be altered slightly without 
changing the overall behaviour.

It is important to ignore this change in the field while modifying the
matrix in the next update step to avoid artifacts (pedestrians moving
backwards without reason).

With this extension, the jam behind a single obstacle composed by a
line of several forbidden cells can be reduced significantly.  The
trade-off lies in an overall reduced flow, as the unhappy pedestrians
gravely disturb the previously unhindered movement of the pedestrians
which pass to the sides of the obstacle.  This evokes the need to find
out for each pedestrian individually whether it might be necessary to
switch to the happy state instantaneously, depending on the local
situation.  This should be achieved without the introduction of
per-pedestrian intelligence.


\section{Simulations}
\label{sec_simus}

In the following we describe the results of simulations of two typical
situations, i.e.\ the evacuation of a large room \cite{panic}
(e.g.\ in the case of a fire) and the formation of lanes in a 
large corridor \cite{social}. We use different
variants of the basic model in order to elucidate the potential of the
different approaches.

\subsection{Evacuation of a large room}
\label{evacuation}

For simplicity, in the case of discrete floor fields pedestrians are
only allowed to move in north ($N$), west ($W$), south ($S$), and east
($E$) direction, which leads to the following form of the 
matrix of preferences:
\begin{equation}
M=\begin{pmatrix}
0 & M_{N} & 0\\
M_{W} & M_{0}& M_{E}\\
0 & M_{S} & 0
\end{pmatrix}
\end{equation}
This choice means no severe restriction since transitions into the
diagonal directions can be implemented quite easily.

In our simulations we have investigated the behaviour of people
leaving a quadratic room with one door only. The $s$-bosonic field has
been chosen such that the occupation number of
$s$-bosons decreases radially
from a maximum value at the door to zero at the corners opposite to
the door. 
\begin{figure}[ht]
\epsfig{file=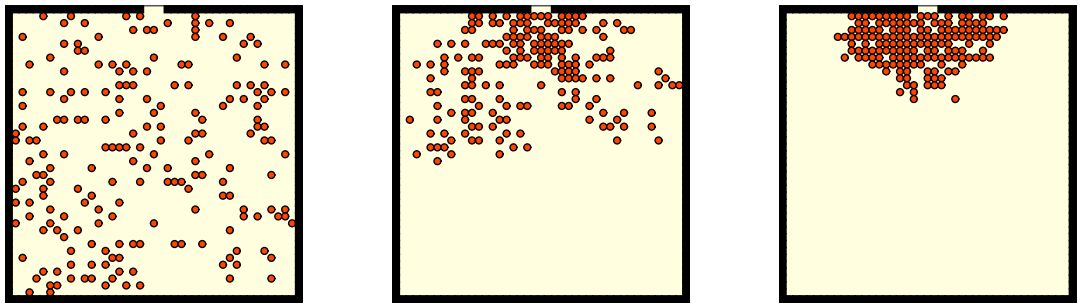}
\caption{People leaving a room with one door only. Displayed are three
  typical stages of the dynamics.}
\label{fig:snaps}
\end{figure}
Typical stages of the dynamics are shown in
Fig.~\ref{fig:snaps}.

\begin{figure}[ht]
\begin{center}
\includegraphics[height=6cm]{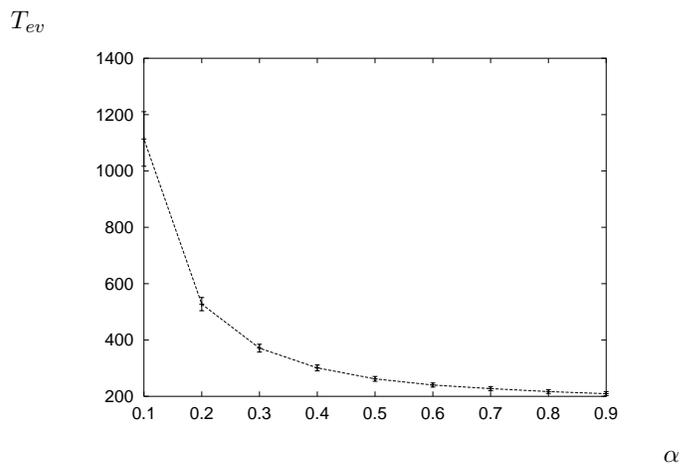}
\end{center}
\caption{Mean evacuation times $T_{ev}$ as function of the decay
  probability $\alpha$ of the $d$-bosons. The 
  parameters are: $J_s=2$, $J_d=J_0=1$, $\beta=10$. The errorbars display 
  the mean standard deviation.}
\label{fig:xeq2.0}
\end{figure}

\begin{figure}[ht]
\begin{center}
\includegraphics[height=6cm]{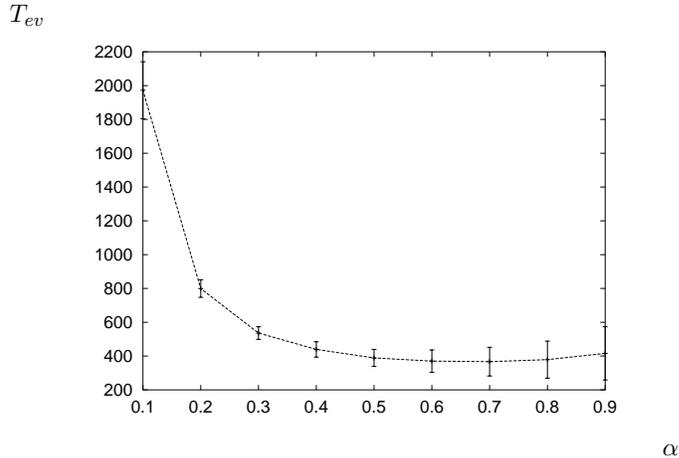}
\end{center}
\caption{Same as Fig.~\ref{fig:xeq2.0}, but with $J_s=1/2$. All other 
parameters are the same.}
\label{fig:xeq0.5}
\end{figure}

As an example we have studied the influence of the lifetime of
$d$-bosons (i.e.\ their decay probability $\alpha$) on the evacuation 
time, i.e., the time it takes for all
people to leave the room. We have seen that if the
coupling strength $J_s$ to the static bosons is rather large,
the evacuation time increases with an increase of the lifetime of
$d$-bosons (see Fig.~\ref{fig:xeq2.0}). Most interestingly, if $J_s$ 
becomes smaller, the best evacuation times are found when the lifetime 
of $d$-bosons is fixed at some intermediate values (see
Fig.~\ref{fig:xeq0.5}).

This finding is very interesting, because it has the consequence that
the attractive interaction of particles can lead to crucial
differences in the particles' behaviour. These changes become more severe if
the particles have no clear idea what the best way to the next exit
is. In addition, one can see that  fluctuations become much more
dominant if $J_s$ goes to zero. Therefore, in case of evacuation
simulations, studying average evacuation times or -- even worse --
looking at one sample only might lead to wrong conclusions. 


\subsection{Lane formation in a long corridor}
\label{laneformation}

We present simulations of a rectangular corridor which is populated by
two species of pedestrians moving in opposite directions. Parallel
to the direction of motion we assume the existence of walls.
Orthogonal to the direction of motion we
investigated both periodic and open boundary conditions.  The length
of the corridor is set to 200 cells.  Widths of 15 and 25 cells have
been used.

With periodic boundary conditions, the density of pedestrians is fixed
for each run.  The program ensures that the overall number of
pedestrians is evenly divided by the numbers for the different species.  For
open boundaries, wie fix the rate at which pedestrians enter the system
at the boundaries (ASEP style insertion rates).  The pedestrians leave
the system as soon as they reach the opposite end of the corridor.

This model clearly provides the option for a complete jam.  The
jamming probability with periodic boundaries at constant density
increases with the length of the system.  An open system can be
thought of as the limit of an infinitely long periodic system,
although density and entry rates do not correspond absolutely (the
density in the open system is always higher than twice the insertion
rate).

The update rules have the same structure as described in Sec.~\ref{sec_floor}.
Only step (6) is modified to

\begin{description}
\item{(6')} The pedestrians change their mode if necessary based on their
  history and the floor field.
They alter the (dynamic) floor field of the cell they occupied before the move.
\end{description}

We performed several runs for different densities and insertion rates,
respectively.  The focus of our attention is the parameter range where
the transition from a stable flow to a complete jam takes place.  The
complete set of parameters for the simulations can be found in
\appref{VoP}.

\onefig{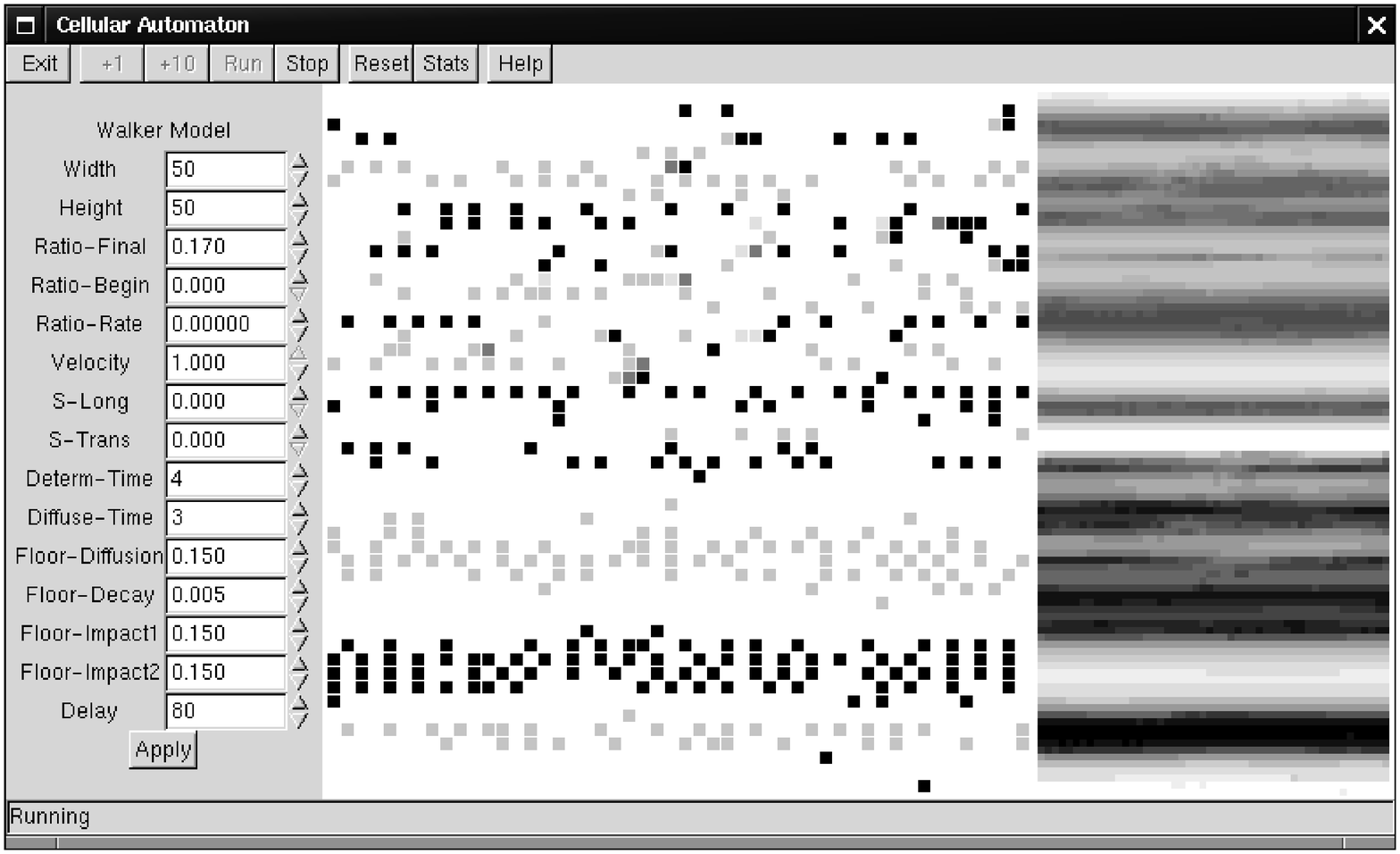}{7cm}
{Snapshot of a simulation with $\rho = 0.17$, $w = h = 50$. The
left part shows the parameter control. The central window is the 
corridor and the light and dark squares are right- and left-moving
pedestrians, respectively. The right part shows the floor fields
for the two species.}{snap1}

\Figref{snap1} shows the graphical frontend running a simulation of
a small periodic system.  The lanes can be spotted easily, both in the
main window showing the cell contents and the small windows on the
right showing the floor field intensity for the two species.

\onefig{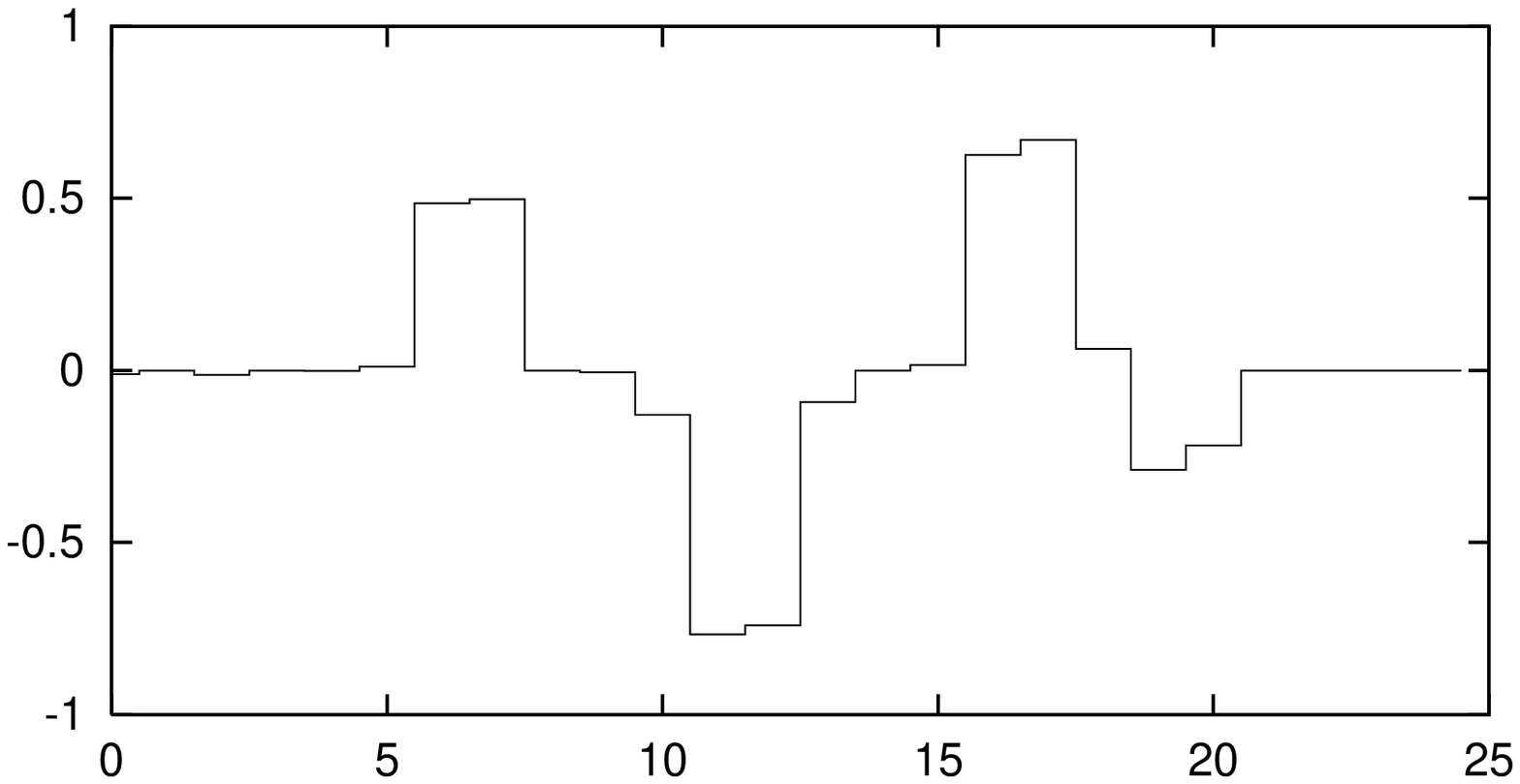}{5cm}
{Velocity profile of a periodic system with $\rho = 0.10$.}{H25R010}

\onefig{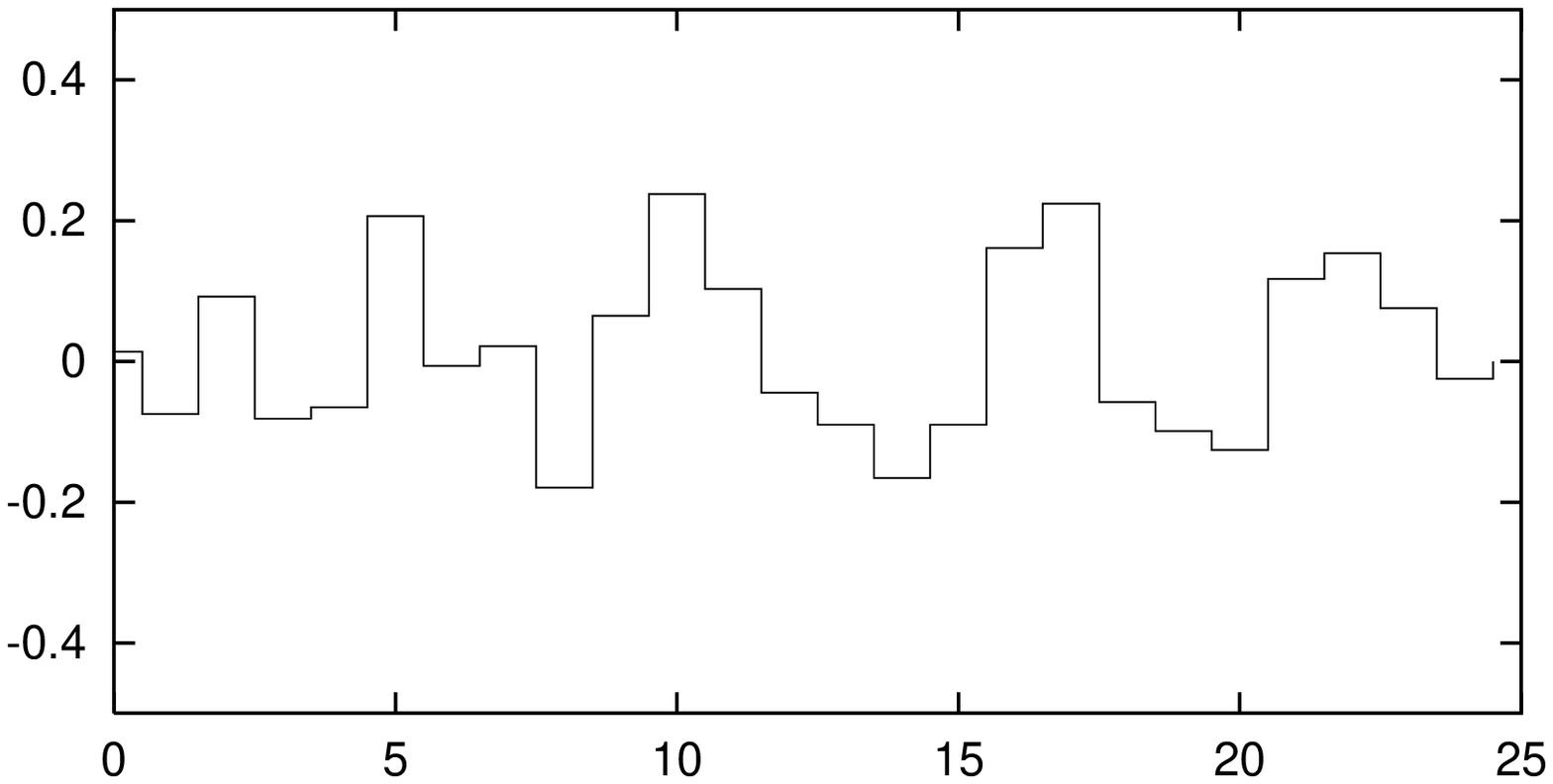}{5cm}
{Velocity profile of an open system taken at $x = \frac{L}{2}$ with
  $\alpha = 0.04$.}{H25A004R0}

\onefig{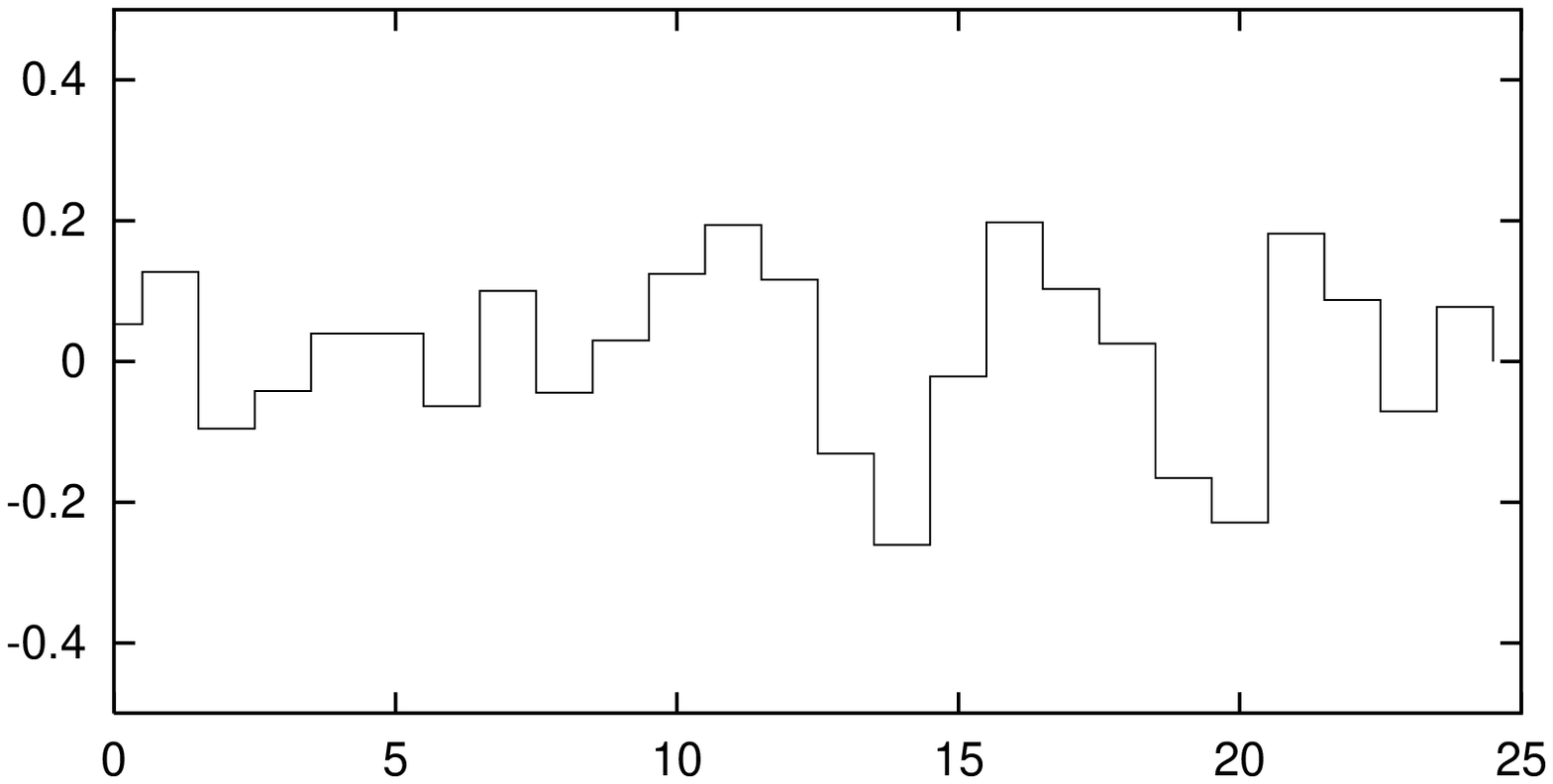}{5cm}
{Velocity profile of an open system taken at $x = \frac{L}{4}$ with
  $\alpha = 0.04$.}{H25A004R1}

To obtain information about the lanes we accumulated the pedestrian
velocities at a cross section perpendicular to the direction of flow.
This is done according to the formula $j_{n+1} = j_{n} \cdot r + v$,
where $j$ is the accumulated value and $v=0,1$ is the velocity of the
pedestrian crossing the line.  $r < 1$ is set to
such a value that the characteristic number of contributing pedestrians
is 100.  Selected profiles are shown in 
Figs.\ \ref{H25R010}--\ref{H25A004R1}.
The values of the other parameters are given in the table in
Appendix \ref{VoP}. Qualitatively our results are in good agreement
with those of \cite{optself} where the lane formation has been
interpreted as an optimal self-organization process.

It is obvious that the lane formation in the periodic system works far
better than in the open system.  The floor field leads to an effective
attraction of identical pedestrians while different pedestrian species
separate.  This results in the formation of a stable pattern in the
periodic case.

In a certain density regime, these lanes are metastable.  Spontaneous
fluctuations can disrupt the flow in one lane causing the pedestrians to
spread and interfere with other lanes.  Eventually the system can run
into a jam by this mechanism.  The average time after which the system
is blocked by a jam is an interesting observable which depends on the
density of pedestrians.
We observe large fluctuations of this observable which require many
samples to find statistically significant information.  

We have also found the formation of an odd number of lanes under
certain conditions. This corresponds to a spontaneous breaking of
the left-right symmetry of the system.

Due to the complexity of our model, the computational speed is
significantly lower compared to the original models of traffic flow.
A typical value measured on a SUN Sparc-10 workstation is $0.24$ mega
updates per second.  It should still be way faster than continuous
models.

\section{Conclusions}

We have introduced a stochastic cellular automaton to simulate
pedestrian behaviour.  We focused on the general concept and the
effects which can be observed with the basic approach, \ie particle
attraction and repulsion between identical and different particles
respectively and lane formation.

The key mechanism is the introduction of the floor field which acts as
a substitute for pedestrian intelligence and leads to collective
phenomena. This floor field makes it possible to translate spatial
long-ranged interactions into non-local interactions in time. The
latter can be implemented much more efficiently on a computer. Another
advantage is an easier treatment of complex geometries. In models with
long-range interactions, e.g.\ the social-force models, one always has
to check explicitly whether pedestrians are separated by walls in
which case there should be no interaction between them. Furthermore,
the computational effort in these models increases proportionally to
the square of the number of individuals.  In contrast, in our approach
it increases only linearly with the system size which is usually
fixed.

The general idea in our model is similar to chemotaxis. However,
the pedestrians leave a virtual trace rather than a chemical one.
This virtual trace has its own dynamics (diffusion and decay) which
e.g.\ restricts the interaction range (in time). It is
realized through a dynamcical floor field which allows to give the
pedestrians only minimal intelligence and to use local interactions.
Together with the static floor field it offers the possibility
to take different effects into account in a unified way, e.g.\
the social forces between the pedestrians or the geometry of the
building.

We presented a rather general form of the model.  Not all the features
are needed in all the cases. E.g. for lane formation we do not need a
static field.  A static field might lead to a "pinning" of lanes.  We
have shown that our model is a good starting point for realistic
applications since it is able to reproduce the basic phenomena
encountered empirically.  In contrast the other CA models so far have
not been shown to exhibit some of the collective phenomena, e.g. lane
formation etc. Other features, e.g.\ oscillations at doors
\cite{social}, have also been observed in our simulations
\cite{cbdiplom}. Quantitative results will presented elsewhere.  The
model can also be applied to more complex geometries and various
characteristics of a crowd can be simulated without major changes. So
it should be possible to study the effects of panic (see \cite{panic}
and references therein).

The description of pedestrians using a cellular automaton approach
allows for very high simulation speeds. Therefore, we have the
possibility to extract the complete statistical properties of our
model using Monte Carlo simulations. This knowldege is of major
importance if one wants to establish risk management techniques that
are nowadays used for the hedging of financial assets all over the
world \cite{bp}.


\section*{Acknowledgment}
\noindent
\noindent{\bf Acknowledgements:} 
Part of this work has been performed within the research program of
the SFB 341 (K\"oln--Aachen--J\"ulich). 
We like to thank H.\ Kl\"upfel, A.\ Ke{\ss}el and D.\ Helbing
for useful discussions.


\appendix

\section{Construction of the matrix of preferences}
\label{MoD}

The aim of this appendix is to show that the matrix of preferences 
can be directly related to observable quantities, namely the average
velocities and their fluctuations. 
The procedure explained here to construct the matrix of preferences is not
essential for the model.  One could freely choose the nine
matrix elements to achieve the desired behaviour of the pedestrians.
However, it is not straightforward to choose five independent
probabilities (one is determined by normalization and three by symmetry) in
a consistent way.  It is therefore convenient to look for a simpler
principle, which might even simplify some calculations.

\begin{figure}[ht]
  \centerline{
    \epsfig{figure=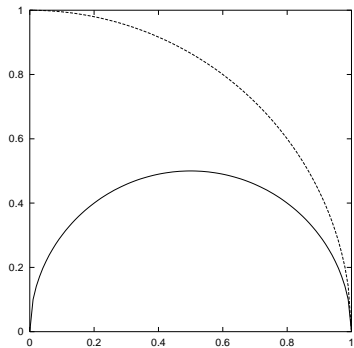,height=5cm}
    }
\caption{$\sigma_l$ and $\sigma_h$ depending on $v$.}
\label{circles}
\end{figure}

We consider first a one-dimensional setup of three adjacent fields which
represents the velocities in $I := \lbrace -1, 0, 1 \rbrace$ from left
to right.  To these cells the probabilities $p_{-1}$, $p_0$ and $p_1$
are assigned.  The values of the average velocity $v$ and of the
standard deviation $\sigma$ are the parameters of this construction.
This leads to three conditions which uniquely determine the probabilities:
\begin{gather}
  \sum_{i \in I} p_i = 1 \\
  \sum_{i \in I} i p_i = v \\
  \sum_{i \in I} (i - v)^2 p_i = \sigma^2
\end{gather}
Not all combinations of $v \in \closediv{-1, 1}$ and $\sigma \in
\closediv{0, 1}$ are allowed.  One finds
\begin{align}
  p_{-1} &= \frac{1}{2} \left( \sigma^2 + v^2 - v \right) \\
  p_0 &= 1 - \left( \sigma^2 + v^2 \right) \\
  p_1 &=\frac{1}{2} \left( \sigma^2 + v^2 + v \right) \, \text{,}
\end{align}
where $\sigma$ is confined to the interval $\closediv{\sigma_l,
\sigma_h}$ with
\begin{align}
  \sigma_l^2 &= \frac{1}{4} - \left( \left \lvert v \right \rvert -
    \frac{1}{2} \right)^2 \\ \sigma_h^2 &= 1 - v^2 \, \text{.}
\end{align}
These restrictions are shown in \figref{circles}.
  
To create a $3 \times 3$ matrix $M$, one creates two such sets:
$\theset{p_{-1}, p_0, p_1}$ with the parameters $\theset{v, \sigma_v}$
which correspond to the forward and backward movement and
$\theset{q_{-1}, q_0, q_1}$ with the parameters $\theset{0, \sigma_t}$
for the (symmetric) transversal movement.  These are simply
multiplied:
\begin{equation}
  M_{ij} = q_i \cdot p_j \, \text{,}
\end{equation}
which produces a movement to the right for positive $v$.  This matrix
is normalized by construction.  We have now achieved a reduction of
free parameters from 5 to 3 and as a side effect formulated a starting
point for analytical calculations.


\section{Typical parameter values}
\label{VoP}

The following table contains the typical parameter values used in
the simulations of lane formation in Sec.~\ref{laneformation}.
\begin{center}
\begin{tabular}{l|c|r}
  Description & Symbol & Value \\
  \hline
  steps to happy transition && 4 \\
  steps to unhappy transition && 3 \\
  floor field diffusion & $D$ & 0.01875 \\
  floor field decay & $\delta$ & 0.005 \\
  first active floor parameter & $b_1$ & 0.15 \\
  second active floor parameter & $b_2$ & 0.15 \\
  first passive floor parameter & $g_1$ & 0.23 \\
  second passive floor parameter & $g_2$ & 0.10
\end{tabular}
\end{center}

The first two lines show the number of consecutive allowed moves which a
particle in unhappy mode needs to become happy again and the
number of consecutive forbidden moves for the inverse transition,
respectively.

The following two lines give the parameters for the modification of the
floor field as shown in (\ref{eq_diffu}).

The active floor parameters determine the influence of the floor field
on the particles (\ref{eq_pij}), whereas the passive floor parameters
describe the action of the particles on the floor field (\ref{eq_g1g2}).

The first six of these values can be found in the lower half of the
configuration panel in \Figref{snap1} (the diffusion constant is
scaled by $\frac{1}{8}$).



\end{document}